\documentstyle[aps,eqsecnum,preprint,epsf]{revtex}
\begin{document}
\draft
\preprint{WIS -- 97/7/Feb -- PH}
\title{Rate equations for quantum transport in multi-dot systems}
\author{S.A. Gurvitz}
\address{Department of Particle Physics, Weizmann Institute of
         Science, Rehovot 76100, Israel}
\date{\today}
\maketitle
\begin{abstract}
Starting with the many-body Schr\"odinger equation we 
derive new rate equations for resonant transport 
in quantum dots linked by ballistic channels with high density of 
states. The charging and the Pauli exclusion principle effects were 
taken into account. It is shown that the current in such a system 
displays quantum coherence effects, even if the dots are away one 
from another. A comparative analysis of  quantum coherence effects  
in coupled  and separated dots is presented. The rate equations are 
extended for description of coherent and incoherent transport 
in arbitrary  multi-dot systems. It is demonstrated that new rate 
equations constitute a generalization of the well-known optical 
Bloch equations. 
\end{abstract}
\section{Introduction}
Quantum transport in small tunneling structures (quantum dots)
have attracted great attention due to the possibility of 
investigating single-electron effects in the electric current\cite{likh}.
Until now research has been mostly concentrated on single dots, but 
the rapid progress in microfabrication technology has made it possible the
extension to coupled dot systems with aligned levels\cite{haug,vdr,wbm}.
An example of such a system is shown schematically in Fig. 1, where
the coupled quantum dot (coupled wells) are  
connected with two separate reservoirs (the leads).
In contrast with a single dot, the electron wave function inside
a coupled dot structure is a superposition of electron 
states localized in each of the dots. As a result, the effects of 
quantum coherence would appear in electron current flowing through 
such a system. Usually these effects are treated in the framework 
of single electron approach. Although this approach is an appropriate 
one for coupled-well semiconductor heterostructures \cite{bry,g,kal}, 
it cannot be applied for coupled dots. The reason is 
the dominant role of the Coulomb interaction that leads to 
the Hubbard-type Hamiltonian for a description of 
these systems\cite{sarma}.
\vskip1cm 
\begin{minipage}{13cm}
\begin{center}
\leavevmode
\epsfxsize=13cm
\epsffile{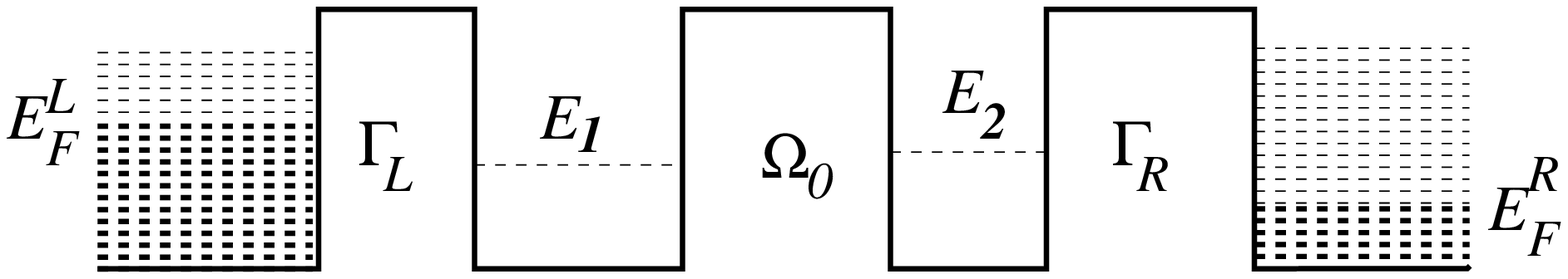}
\end{center}
{\begin{small}
Fig.~1. Resonant transport through a coupled-dot structure.
$\Gamma_{L,R}$ denote the corresponding tunneling rate from (to) the 
left (right) reservoirs, and $\Omega_0$ is the interdot hopping 
amplitude. $E_F^{L,R}$ denote the Fermi level in the left (right)
reservoirs. Only those energy levels inside the dots that carry 
electron current are shown. 
\end{small}}
\end{minipage} \\ \\ 
  
The Coulomb blockade effects in electron transport through 
a {\em single} dot can be taken into account in the most simple way by 
using the ``classical'' rate equations\cite{glaz,ak,been,Davies4603}.
These equations describe the current through the dot 
in terms of balance between incoming and outgoing rates 
from (to) the leads. Basically the same classical 
approach has been used to calculate the conductance
of coupled multi-dot systems\cite{klim,chen}. In these works the 
whole array of quantum dots has been treated as a single quantum 
system, where its many-body eigenstates were calculated by exact (numerical) 
diagonalization of the Hubbard Hamiltonian. Then the leads (reservoirs)
were incorporated through the rate equations for single dots\cite{been}.
Yet, such a procedure is correct only for small coupling with 
the leads. If the corresponding rates $\Gamma$ are of the order 
of the interdot transitions $\Omega$ (Fig. 1), they affect
the diagonalization of the Hubbard Hamiltonian\cite{g}.
Besides, the approach\cite{klim,chen} is mainly numerical
and the inelastic scattering within the array is accounted
for phenomenologically.  
 
A different way to treat the quantum transport in multi-dot 
systems is to use the Bloch-type rate 
equations\cite{naz,glp1,gp,g1} instead of the classical rate equations.
In contrast with previous treatments\cite{klim,chen} 
this approach is valid also for strong 
coupling with the leads and it accounts the inelastic 
scattering effects inside the multi-dot system\cite{gp}. 
It proves to be much simpler than the other approaches,
so that in many cases the analytical treatment of the problem is 
possible\cite{glp1,gp}. It is also important to note that this approach 
is not a phenomenological one: new generalized Bloch-type rate equations  
were {\em derived} from the many-body Schr\"odinger equation 
with the Hubbard Hamiltonian by integrating out the continuum 
states of the reservoirs\cite{gp}.

An important advantage of the Bloch-type equations is a clear
distinction between coherent(``quantum'') and non-coherent 
(``classical'') terms.   
The quantum coherence in the Bloch equations manifests itself 
in the non-diagonal density-matrix elements (``coherences''), coupled 
with the diagonal density-matrix elements 
(``probabilities'')\cite{bloch}. 
Such a coupling always takes place, whenever a carrier 
jumps between {\em isolated} states inside the system\cite{gp}. 
Otherwise, the diagonal and non-diagonal matrix elements 
are decoupled and the Bloch equations turn into 
the {\em classical} rate equation. 

Actually Bloch equations describe quantum motion between 
isolated (non-orthogonal) states, which are directly coupled
(as in the double-well potential). In general, these states can be   
separated by a medium with high density of states.
An example of such a system is shown in Fig. 2, where two dots 
with aligned levels are separated by the 
ballistic channel. In this case  
a carrier transport from the left to the right reservoir
proceeds through continuum states in the 
channel. Usually transitions from discrete to
continuum states lead to dephasing, i.e. to destruction 
of quantum coherence. Yet, some coherent effects may survive. 
It is therefore important to establish of whether quantum coherence 
still affects the transport through separated dots, or
more precisely: are the diagonal 
density-matrix elements still coupled with the non-diagonal matrix 
elements in the corresponding rate equations? 
\vskip1cm
\begin{minipage}{13cm}
\begin{center}
\leavevmode
\epsfxsize=13cm
\epsffile{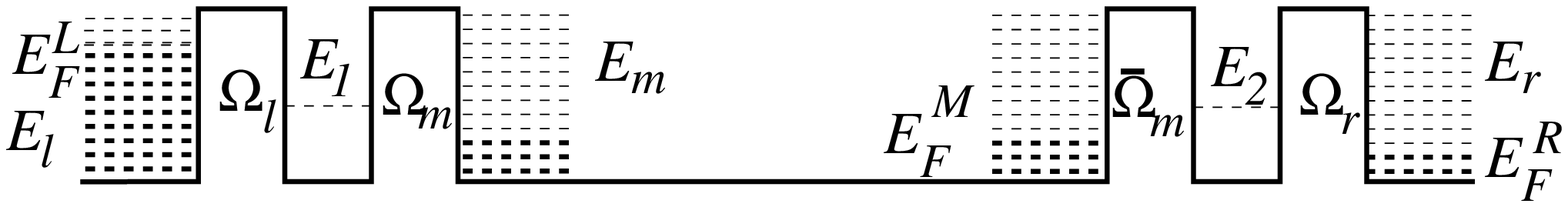}
\end{center}
{\begin{small}
Fig.~2. Resonant transport through two quantum dots separated  
by a ballistic channel. Here $\Omega_l$ and $\Omega_r$ denote the coupling of 
the left and right dots dot with the levels $E_l$ and $E_r$ in the 
left and in the right leads. $\Omega_m$ and $\bar\Omega_m$ denote 
the coupling of the left and the right dots with the level $E_m$ in 
the ballistic channel. 
\end{small}}
\end{minipage} \\ \\ 

This question leads us to a more general problem, of how to modify 
the Bloch rate equations for transport in coupled-dots for a 
general case of arbitrary distributed quantum dots. Similar to  
the Bloch equations one can anticipate that that the new rate 
equations would incorporate the classical and
quantum descriptions in a very efficient way.
Therefore these equations can provide most useful account of  
quantum coherence effects in different processes 
in a comparison with usual approaches\cite{rammer}. 

In this paper we study the above mentioned problems 
taking as a basis the microscopic many-body Schr\"odinger equation with 
the Hubbard-type tunneling Hamiltonian, describing the entire system of the 
reservoirs and the quantum dots.  
The plan of the paper is the following. In Sect. 2 we describe 
the previously derived Bloch-type rate equations for  
coupled dots and some particular features of the coherent transport 
in these systems.  In Sect. 3 we present the rate equations 
for separated dots and outline their  
derivation from the microscopic Schr\"odinger equation. 
The details of the derivation are described in Appendix. 
A comparison of coherent effects in quantum transport 
through coupled and separated dots is given in Sect. 4. The quantum rate
equations for a general configuration of quantum dots are  
presented in Sect. 5. The last section is a summary.

\section{Rate equations and coherent effects in coupled quantum-dots}

We start with a review of the quantum rate equations for coupled multi-dot
systems, connected with two reservoirs 
(the ``emitter'' and the ``collector'') and interacting with phonon reservoir. 
The entire system is described by the Hubbard-type tunneling Hamiltonian 
where the electron-electron interaction is taken into account  
by introducing the corresponding electrostatic charging 
energy. In case of large voltage bias one can reduce 
the many-body Schr\"odinger equation to the system of quantum rate equations 
by integrating out continuum reservoir states\cite{gp}. As a result 
the following Bloch-type equations for the density matrix of 
the multi-dot system $\{\sigma_{\alpha\beta}\}$ are obtained, 
where $\alpha ,\beta$ denote the 
isolated (non-orthogonal) states of the system:   
\begin{mathletters}
\label{c3}
\begin{eqnarray}
\dot\sigma_{\alpha\alpha} & = &
 i\sum_{\beta}\Omega_{\alpha\to\beta}
(\sigma_{\alpha\beta}-\sigma_{\beta\alpha})
-\sigma_{\alpha\alpha} \sum_{\gamma}
\Gamma_{\alpha \rightarrow \gamma} +
 \sum_{\delta} \sigma_{\delta\delta}
\Gamma_{\delta \rightarrow \alpha}\;,
 \label{c3a}\\
\dot\sigma_{\alpha\beta} & = & i(E_\beta - E_\alpha) \sigma_{\alpha\beta} +
i\left (\sum_{\gamma}\sigma_{\alpha\gamma}\Omega_{\gamma\to\beta}
-\sum_{\delta}\Omega_{\alpha\to\delta}
\sigma_{\delta\beta}\right )
\nonumber\\
    &  &~~~~~~~~~~~~~~~~~~~~~~~~~~~~ -\frac{1}{2}\sigma_{\alpha\beta}
   \sum_{\delta}\left (\Gamma_{\alpha \rightarrow\delta}
    +\Gamma_{\beta \rightarrow \delta}\right )
                  +\sum_{\gamma\delta\neq \alpha\beta}
\sigma_{\gamma\delta}
        \Gamma_{\gamma\delta\rightarrow \alpha\beta}.
\label{c3b}
\end{eqnarray}
\end{mathletters}
Here $\Omega_{\alpha\to\beta}$ is the amplitude of one-electron hopping 
that results in the transition between the states $\alpha$ and $\beta$.
The width $\Gamma_{\alpha\to \gamma}=2\pi\rho |\Omega_{\alpha\to \gamma} |^2$ 
is the probability per unit time for the system to make a transition from
the state $|\alpha\rangle$ to
the state $|\gamma\rangle$ of the device due to the tunneling to (or from) the
reservoirs, or due to interaction with the phonon bath, or any other 
interaction generated by a continuum state medium with the density of states 
$\rho$. Notice that Eq.~(\ref{c3a}) for diagonal elements has 
a form of classical rate equations, 
except for the first term. This term is generated by an electron hopping
between the {\em isolated} levels, which results in the coupling  
with non-diagonal density-matrix elements. Therefore it
is responsible for coherent quantum effects in electron transport.
The nondiagonal matrix elements are described by Eq.~(\ref{c3b}).
The last term in Eq.~(\ref{c3b}) appears 
only for systems with the number of isolated 
states participating in the transport is more than two. 
It describes the simultaneous conversion of the states 
$\gamma\to\alpha$ and $\delta\to\beta$,  
generated by the same one-electron decay to (from) continuum. 

The current flowing through the system is given by  
\begin{equation}  
I(t)=\sum_\gamma\sigma_{\gamma\gamma}(t)\Gamma^{(\gamma)}_R, 
\label{c4}
\end{equation}
where the sum is extended over states $|\gamma\rangle$ in which the dot 
adjacent to the collector is occupied (we consider the electron charge $e=1$). 
$\Gamma_R^{(\gamma)}$ is the partial width of the state $|\gamma\rangle$
due to tunneling to the collector (the right reservoir). 

It follows from Eqs.~(\ref{c3}) that the coherent effects 
do appear in the quantum transport whenever a carrier 
jumps from one to another {\em isolated} state inside the device. 
In the absence of such transition as, for
instance, in resonant tunneling through a single dot, 
the diagonal and non-diagonal matrix elements 
are {\em decoupled} and the evolution of diagonal density-matrix elements is 
described by the {\em classical} rate equation.

For an example we consider the quantum transport through 
a double-dot system at zero temperature\cite{foot1}, 
shown in Fig. 1. In order to diminish the number of equations  
we assume that the Coulomb repulsion does not allow for  
two electrons to occupy the same dot, i.e. $E_i+U_{ii}\gg E_F^L$, 
where $i=1,2$ and $U_{ii}$ is the corresponding 
charging energy. Yet, the  
interdot Coulomb repulsion $U_{12}$ is much smaller, so the 
both levels $E_{1,2}$ can be occupied simultaneously.  
In this case there are four available states of the double-dot system
(for simplicity we neglected the spin):
$|a\rangle$ -- the levels $E_{1,2}$ are empty,
$|b\rangle$ -- the level $E_1$ is occupied, 
$|c\rangle$ -- the level $E_2$ is occupied,
$|d\rangle$ -- the both level $E_{1,2}$ are occupied.
Using Eqs.~(\ref{c3}) with $\alpha ,\beta,\ldots =\{ a,b,c,d\}$ 
we find the following equations describing the time evolution 
of the corresponding density-matrix elements
\begin{mathletters}
\label{c1}
\begin{eqnarray}
\dot\sigma_{aa} & = & -\Gamma_L\sigma_{aa}
+\Gamma_R\sigma_{cc}\;,
\label{c1a}\\
\dot\sigma_{bb} & = & \Gamma_L\sigma_{aa}
+\Gamma'_R\sigma_{dd}+i\Omega_0(\sigma_{bc}-\sigma_{cb})\;,
\label{c1b}\\
\dot\sigma_{cc} & = & -\Gamma_R\sigma_{cc}
-\Gamma'_L\sigma_{cc}-i\Omega_0(\sigma_{bc}-\sigma_{cb})\;,
\label{c1c}\\
\dot\sigma_{dd} & = & -\Gamma'_R\sigma_{dd}
+\Gamma'_L\sigma_{cc}\;,
\label{c1d}\\
\dot\sigma_{bc} & = & i(E_2-E_1)\sigma_{bc}+
i\Omega_0(\sigma_{bb}-\sigma_{cc})
-\frac{1}{2}(\Gamma'_L+\Gamma_R)\sigma_{bc}\;,
\label{c1e}
\end{eqnarray}
\end{mathletters}
which are supplemented with the probability conservation condition:
$\sum_i\sigma_{ii}(t)=1$. Note that the interdot Coulomb repulsion results in a variation 
of the corresponding transition rate whenever 
both levels are occupied  ($\Gamma_{L,R}$ is replaced by $\Gamma'_{L,R}$,
evaluated at the energy $E_{1,2}+U_{12}$ respectively\cite{gp}).

The dc current is 
$I= \Gamma_R\sigma_{cc}(t\to\infty )+ \Gamma'_R \sigma_{dd}(t\to\infty )$, 
Eq.~(\ref{c4}).  In order to shorten the final expression 
we assume that the rates are weakly dependent on the energy, 
$\Gamma'_{L,R}=\Gamma_{L,R}$.
Then solving Eqs.~(\ref{c1}) in the limit $t\to\infty$ we obtain 
\begin{equation}
I=\left (\frac{\Gamma_L\Gamma_R}{\Gamma_L+\Gamma_R}\right )
\frac{\Omega_0^2}{\epsilon^2\Gamma_L\Gamma_R/
(\Gamma_L+\Gamma_R)^2+\Omega_0^2+\Gamma_L\Gamma_R/4},
\label{c2}
\end{equation} 
where $\epsilon =E_1-E_2$.

It is interesting to compare Eqs.~(\ref{c1}) with the corresponding 
classical rate equations where only transitions between 
the {\em diagonal} density-matrix elements take place. 
In this case the probabilities $\sigma_{bb}$ and $\sigma_{cc}$
of finding an electron in the first and the second dot are 
coupled by the some rate $\Gamma_M$.
One easily finds that Eqs.~(\ref{c1}) become 
\begin{mathletters}
\label{c5}
\begin{eqnarray}
\dot\sigma_{aa} & = & -\Gamma_L\sigma_{aa}
+\Gamma_R\sigma_{cc}\;,
\label{c5a}\\
\dot\sigma_{bb} & = & \Gamma_L\sigma_{aa}
+\Gamma_R\sigma_{dd}-\Gamma_M(\sigma_{bb}-\sigma_{cc})\;,
\label{c5b}\\
\dot\sigma_{cc} & = & -\Gamma_R\sigma_{cc} 
-\Gamma_L\sigma_{cc}+\Gamma_M(\sigma_{bb}-\sigma_{cc})\;,
\label{c5c}\\
\dot\sigma_{dd} & = & -\Gamma_R\sigma_{dd}
+\Gamma_L\sigma_{cc}\;.
\label{c5d}
\end{eqnarray}
\end{mathletters}
(Cf. with the analogues rate equations in\cite{bjorn,gm}).
Solving these classical rate equations in the limit $t\to\infty$ we 
obtain for the (classical) dc current
\begin{equation}
I_{cl}=\frac{\Gamma_L\Gamma_R\Gamma_M}
{\Gamma_L\Gamma_R+\Gamma_L\Gamma_M+\Gamma_R\Gamma_M}.
\label{c6}
\end{equation}   

Comparing Eq.~({\ref{c6}) with its ``quantum'' counterpart, 
Eq.~(\ref{c2}), we find that the quantum-mechanical nature of the 
dc current displays itself in 
the Lorentzian-shape resonance as a function 
of the levels disalignment, $\epsilon =E_2-E_1$. 
Such resonances were already observed in
the coupled-dot systems\cite{vdr}. Yet, the existence of the resonance 
is not necessarily related to the quantum coherence effect. The later 
manifests itself in a peculiar and even a counter-intuitive 
dependence of the resonant current on coupling with the reservoirs. 
It follows from Eq.~(\ref{c2}) that 
the current {\em decreases} when the coupling with the reservoirs 
{\em increases}. For instance, the peak value of the resonant current 
$I_{\epsilon=0}\propto \Gamma_L/\Gamma_R\to 0$ for 
$\Gamma_R\gg\Gamma_L$. It implies that by 
{\em increasing} the penetrability of the barrier, connecting the second 
dot with the collector we {\em diminish} the total current.
As an example we show in Fig.~3 the peak value of the resonant current
as a function of $\Gamma_R$ for constant $\Gamma_L$ and $\Omega_0$
(the solid line). This effect can be observed in coupled-dot structures by 
changing the corresponding gate voltage.  
\vskip1cm 
\begin{minipage}{13cm}
\begin{center}
\leavevmode
\epsfxsize=10cm
\epsffile{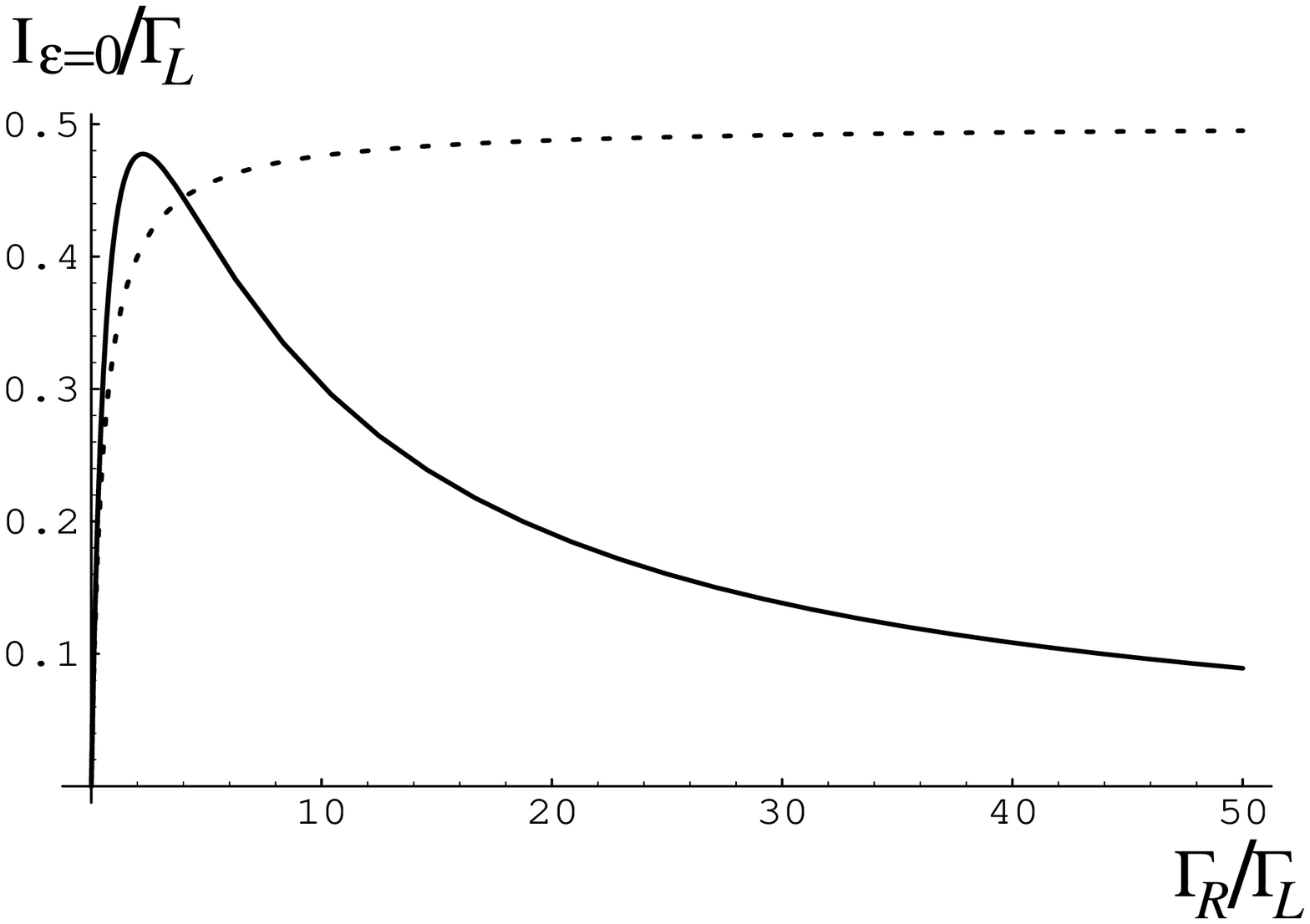}
\end{center}
{\begin{small}
Fig.~3. Current through the coupled-dot structure, Fig. 1, as
a function of $\Gamma_R/\Gamma_L$. 
The solid line corresponds to Eq.~(2.4) for $\epsilon=0$ and 
$\Gamma_L^2/4\Omega_0^2=5$. The dashed line is the result of 
the classical rate equations, Eq.~(2.6), for $\Gamma_M=\Gamma_L$.
\end{small}}
\end{minipage} \\ \\ 
For a comparison the corresponding ``classical'' current 
$I_{cl}$, Eq.~(\ref{c6}), is shown in Fig. 3 by the dashed line. 

The decrease of the resonant current shown in Fig. 3  
is a result of destructive quantum interference during  
decay of a coherent superposition to continuum in the regime 
of strong coupling with the reservoirs\cite{gm,gbj}. A similar effect 
had been observed in electron decay from coupled quantum wells 
to continuum\cite{bg}. It is quite interesting that 
such a counter-intuitive behavior of the resonant current 
looks as a manifestation of the Zeno effect\cite{zeno}.
It tells us that the quantum transitions between different states 
slow down when one of the states is continuously observed. Indeed, 
the increase of the right barrier penetrability leads to
immediate tunneling of an electron to the collector whenever it arrives 
to the second dot.
It can be considered as a continuous observation of the second dot state,
which results in an effective electron localization in the first 
dot\cite{kor}.
  
\section{Rate equations for two separated dots}

Now we are going to derive rate equations for quantum transport 
in mesoscopic systems with arbitrary configuration 
of quantum dots. As a generic example 
we consider quantum transport through two quantum dots separated by 
a ballistic channel, Fig. 2. 
The dots contain only isolated levels, whereas 
the density of states in the ballistic channel and in the emitter and the
detector is very high (continuum).   
This system can be described by the tunneling Hamiltonian 
\begin{eqnarray}
{\cal H} &=&\sum_l E_{l}a^{\dagger}_{l}a_{l} +
E_1 a_1^{\dagger}a_{1}+
\sum_m E_{m}a^{\dagger}_{m}a_{m} +
E_2 a_2^{\dagger}a_{2}\sum_r a_{r}a^{\dagger}_{r}a_{r}
+\sum_{i,j=1,2}U_{ij}n_in_j
\nonumber\\ 
&+& \left\{ \sum_l \Omega_{l}a^{\dagger}_{1}a_l 
+ \sum_m \Omega_{m}a^{\dagger}_ma_{1} 
+ \sum_m \bar\Omega_{m}a^{\dagger}_{2}a_m 
+ \sum_r \Omega_{r}a_r^{\dagger}a_2 + {\rm H.c.}\right\}
\label{a1}
\end{eqnarray}
Here the subscripts $l$, $m$ and
$r$ enumerate correspondingly the levels in the
left reservoir, in the (middle)
ballistic channel and in the right reservoir. 
The spin degrees of freedom were omitted. In order to 
simplify the derivation we assumed that the intradot charging 
energy $U_{ii}$ is large, $E_{1,2}+U_{ii}\gg E_F^L$. Thus  
only one electron can occupy each of the dots.
However, the interdot charging energy $U_{12}$
is much smaller, so it does not prevent simultaneous occupation 
of the two dots. The same is assumed for 
the Coulomb repulsion between electrons inside the dots and 
the ballistic channel. Although we did not include this interaction 
in the Hamiltonian (\ref{a1}), it can be treated in the same way 
as the interdot interaction $U_{12}$. 
As in the previous case we restrict ourselves to the zero temperature case,
even though the results are valid for a finite temperature, as 
would be clear from the derivation.    

Let us assume that all the levels in the emitter, in the 
ballistic channel and in the collector are initially filled up to the Fermi 
energies $E_F^L$, $E_F^M$ and $E_F^R$ respectively. We call 
it as the ``vacuum'' state, $|0\rangle$. (In the following we consider 
the case of large bias, so that $E_F^L\gg E_F^M, E_F^R$).
This vacuum state is unstable: the Hamiltonian
Eq.~(\ref{a1}) requires it to decay exponentially to continuum states 
having the form $a_{1}^{\dagger}a_{l}|0\rangle$ with an electron in
the level $E_1$ and a hole in the emitter continuum, 
$a_{m}^{\dagger}a_{l}|0\rangle$ with an electron in
the level $E_m$ in the ballistic channel and a hole in the emitter, and so on. 
The many-body wave function describing this system can be written 
in the occupation number representation as 
\begin{eqnarray}
|\Psi (t)\rangle & = & \left [ b_0(t) + \sum_l b_{1l}(t)
     a_{1}^{\dagger}a_{l} + \sum_{l,m} b_{lm}(t)a_{m}^{\dagger}a_{l}
   +\sum_l b_{2l}(t)a_{2}^{\dagger}a_{l}  \right.
      \nonumber\\
 &+& \left. \sum_{l,r} b_{lr}(t)a_{r}^{\dagger}a_{l} 
           +\sum_{l<l'} b_{12ll'}(t)a_{1}^{\dagger}a_{2}^{\dagger}a_{l}a_{l'} 
           +\sum_{l<l',r} b_{1ll'r}(t)
           a_{1}^{\dagger}a_{r}^{\dagger}a_{l}a_{l'}
           + \ldots \right ] |0\rangle, 
\label{a2}
\end{eqnarray}
where $b(t)$ are the time-dependent probability amplitudes to
find the system in the corresponding states described above. 
These amplitudes are obtained from the Shr\"odinger equation 
$i|\dot\Psi (t)\rangle ={\cal H}|\Psi (t)\rangle$, supplemented 
with the initial condition  
($b_0(0)=1$, and all the other $b(0)$'s being zeros). Using the amplitudes
$b(t)$ we can find the density-matrix of the quantum dots, 
$\sigma_{ij}^{(k,n)}(t)$, by tracing out the continuum 
states of the reservoirs and the ballistic channel. Here the subscript 
indices in $\sigma$    
denote four states of the dots:  $i,j=\{a,b,c,d\}$, where
$|a\rangle$ -- the levels $E_{1,2}$ are empty,
$|b\rangle$ -- the level $E_1$ is occupied, 
$|c\rangle$ -- the level $E_2$ is occupied,
$|d\rangle$ -- the both level $E_{1,2}$ are occupied, 
and the superscript indices $k,n$ denote the number of 
electrons accumulated in the ballistic channel and in the collector 
respectively at time $t$, Fig. 4. 
\vskip1cm 
\begin{minipage}{13cm}
\begin{center}
\leavevmode
\epsfxsize=11cm
\epsffile{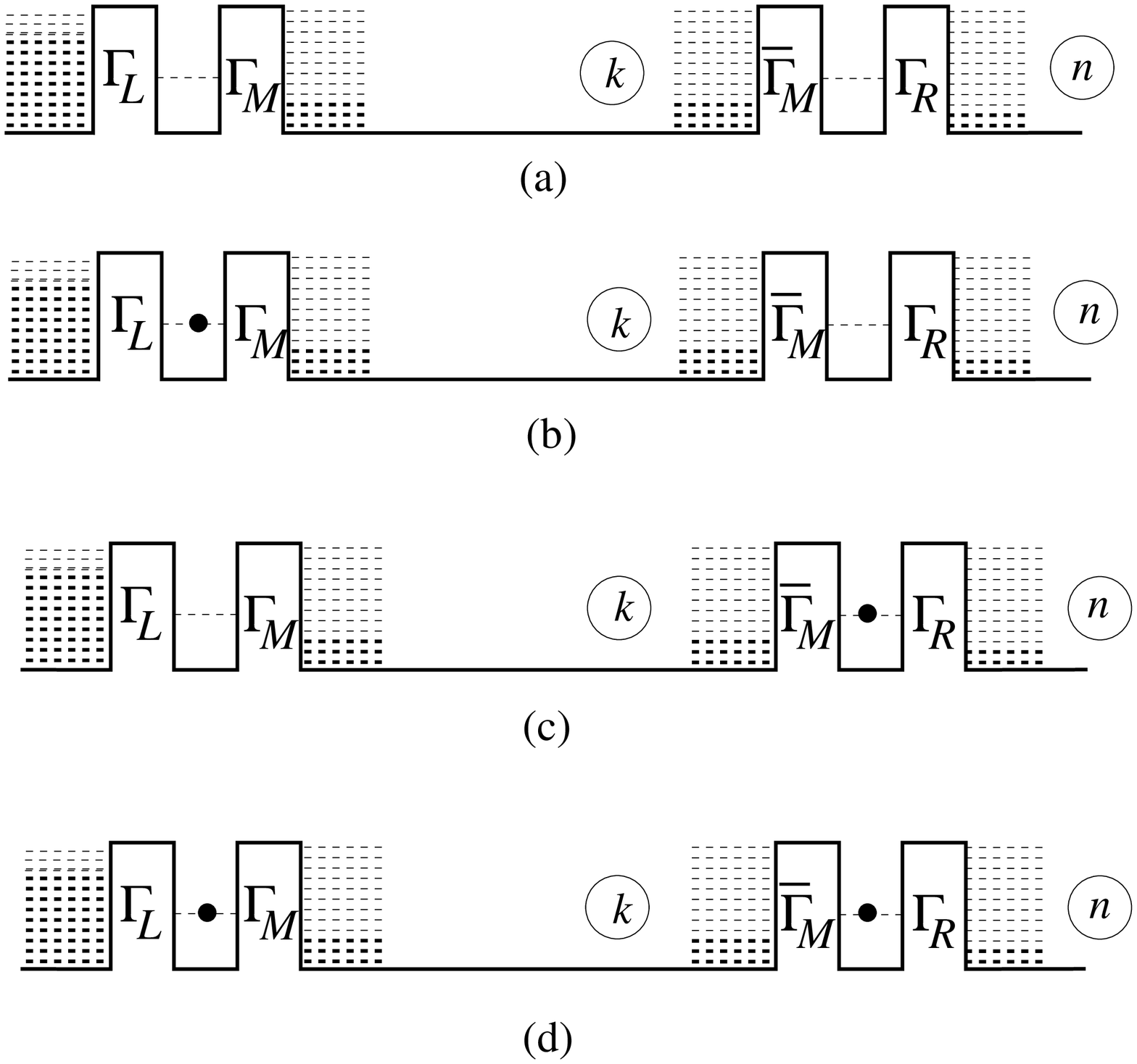}
\end{center}
{\begin{small}
Fig.~4. Electron states of the two separated dot structure, shown in Fig. 2. 
$\Gamma_{L,R}$, $\Gamma_M$ and $\bar\Gamma_M$ are the tunneling rates
between the dots and the reservoirs, and between the dots and the 
ballistic channel. The indices $k$ and $n$ denote the number of electrons 
penetrating to the ballistic channel and to the collector at time $t$. 
\end{small}}
\end{minipage} \\ \\ 
One finds  
\begin{mathletters}
\label{a7}
\begin{eqnarray}
&&\sigma_{aa}^{(0,0)}(t) = |b_{0}(t)|^2,~~~
\sigma_{aa}^{(1,0)}(t)= \sum_{l,m} |b_{lm}(t)|^2,~~~
\sigma_{aa}^{(1,1)}(t)=\sum_{l<l',m,r} |b_{ll'mr}(t)|^2,~~~ \ldots
\label{a7a}\\
&&\sigma_{bb}^{(0,0)}(t) =\sum_l |b_{1l}(t)|^2,~~~
\sigma_{bb}^{(1,0)}(t)=\sum_{l<l',m} |b_{1ll'm}(t)|^2,~~~
\nonumber\\
&&~~~~~~~~~~~~~~~~~~~~~~~~~~~~~~~~~~~~~~~~~~~~~~~~~~~~~
\sigma_{bb}^{(1,1)}(t)=\sum_{l<l'<l'',m,r} |b_{1ll'l''mr}(t)|^2,~~~ \ldots
\label{a7b}\\
&&\cdots\cdots\cdots\cdots\cdots 
\nonumber\\
&&\sigma_{bc}^{(0,0)}(t) = \sum_l b_{1l}(t)b^*_{2l}(t),~~~
\sigma_{bc}^{(1,0)}(t)= \sum_{l<l',m} b_{1ll'm}(t)b^*_{2ll'm}(t),~~~
\nonumber\\
&&~~~~~~~~~~~~~~~~~~~~~~~~~~~~~~~~~~~~~~~~~~~~~~~
\sigma_{bc}^{(1,1)}(t)=\sum_{l<l'<l'',m,r} b_{1ll'l''mr}(t)b^*_{2ll'l''mr}(t),
~~~\ldots 
\label{a7c}
\end{eqnarray}
\end{mathletters}
\par
The rate of electrons arriving to the collector determines the electron 
current in the system. Therefore the current operator is 
$\hat I=i[{\cal H},\hat N_R]$, where 
$\hat N_R=\sum_ra^{\dagger}_ra_r$ is the operator for 
the total number of electrons accumulated in the right reservoir. 
Using Eqs.~(\ref{a1}), (\ref{a2}), (\ref{a7}) we find that the 
current $I(t)$ flowing through the system is 
\begin{equation}
I(t)=\langle\Psi (t)|\hat I|\Psi (t)\rangle =
\sum_{k,n} n\left [\dot\sigma_{aa}^{(k,n)}(t)+
\dot\sigma_{bb}^{(k,n)}(t)+
\dot\sigma_{cc}^{(k,n)}(t)+
\dot\sigma_{dd}^{(k,n)}(t)\right ] 
\label{a8}
\end{equation}
As expected, $I(t)$ is the time derivative of the total charge 
accumulated in the collector.

It follows from Eq.~(\ref{a8}) that the current $I(t)$ flowing 
through this system is expressed in terms of the diagonal  
elements of the density-matrix $\sigma (t)$. 
In order find the differential equations for 
$\sigma (t)$ we need to sum over the states of 
the reservoirs and the ballistic channel, Eqs.~(\ref{a7}). It can be done 
analytically by using the procedure similar to that in Refs.\cite{gp,g1}, 
providing that the levels are not close to Fermi levels,
$|E_i-E_F|\gg\Gamma$, Fig. 4. The details of derivation are 
described in Appendix. 
As a result we obtain the following Bloch-type rate equations for the matrix 
elements of the density-submatrix $\sigma (t)$:
\begin{mathletters}
\label{a13}
\begin{eqnarray}
&&\dot{\sigma}^{(k,n)}_{aa} = - \Gamma_L \sigma_{aa}^{(k,n)}
+\Gamma_M\sigma_{bb}^{(k-1,n)}+\bar\Gamma_M\sigma_{cc}^{(k-1,n)}
+\Gamma_R\sigma_{cc}^{(k,n-1)}\nonumber\\
&&~~~~~~~~~~~~~~~~~~~~~~~~~~~~~~~~~~~~~~~~~~~~~~~~~~~~~~~~~~~~~~
+2\pi\rho_M\Omega_M\bar\Omega_M(\sigma^{(k-1,n)}_{bc}+\sigma^{(k-1,n)}_{cb})
\label{a13a}\\
&&\dot{\sigma}^{(k,n)}_{bb} = - \Gamma_M \sigma_{bb}^{(k,n)}
+\Gamma_L\sigma_{aa}^{(k,n)}+\bar\Gamma_M\sigma_{dd}^{(k-1,n)}
+\Gamma_R\sigma_{dd}^{(k,n-1)}\nonumber\\
&&~~~~~~~~~~~~~~~~~~~~~~~~~~~~~~~~~~~~~~~~~~~~~~~~~~~~~~~~~~~~~~~~~~~~
-\pi\rho_M\Omega_M\bar\Omega_M(\sigma^{(k,n)}_{bc}+\sigma^{(k,n)}_{cb})
\label{a13b}\\
&&\dot{\sigma}^{(k,n)}_{cc} = -(\Gamma_L +\bar\Gamma_M +\Gamma_R) 
\sigma_{cc}^{(k,n)}
+\Gamma_M\sigma_{dd}^{(k-1,n)}
-\pi\rho_M\Omega_M\bar\Omega_M(\sigma^{(k,n)}_{bc}+\sigma^{(k,n)}_{cb})
\label{a13c}\\
&&\dot{\sigma}^{(k,n)}_{dd} = -(\Gamma_M +\bar\Gamma_M +\Gamma_R) 
\sigma_{dd}^{(k,n)}+\Gamma_L\sigma_{cc}^{(k,n)}
\label{a13d}\\
&&\dot{\sigma}^{(k,n)}_{bc} = i(E_2-E_1)\sigma^{(k,n)}_{bc}
-\pi\rho_M\Omega_M\bar\Omega_M(\sigma^{(k,n)}_{bb}+\sigma^{(k,n)}_{cc})
+2\pi\rho_M\Omega_M\bar\Omega_M\sigma_{dd}^{(k-1,n)}\nonumber\\
&&~~~~~~~~~~~~~~~~~~~~~~~~~~~~~~~~~~~~~~~~~~~~~~~~~~~~~~~~~~~~~~~~~~
-\frac{1}{2}(\Gamma_L+\Gamma_M +\bar\Gamma_M +\Gamma_R) 
\sigma_{bc}^{(k,n)}\, .
\label{a13e}
\end{eqnarray}
\end{mathletters}
Here $\Gamma$ denote the tunneling widths to the leads and the ballistic 
channel ($\Gamma =2\pi\rho |\Omega |^2$, 
where $\rho $ is the corresponding density of states), and  
$\Omega_M$, $\bar\Omega_M$ are the hopping 
amplitudes between the left and the right dots and
the states $E_m=E_{1,2}$ of the ballistic channel.

Equations (\ref{a13}) have clear physical interpretation. Consider 
for instance Eq.~(\ref{a13a}) for the probability rate of finding  
the system in the state $a$ with  
$k$ electrons in the ballistic channel and $n$ electrons 
in the right reservoir (Fig. 4a). This state decays with the rate $\Gamma_L$
into the state $b$ (Fig. 4b) whenever an electron enters the first 
dot from the left reservoir. This process is  
described by the first term in Eq.~(\ref{a13a}). On the other hand,
the states $b$ and $c$ (Figs.~4b, 4c) with $k-1$ electrons 
in the ballistic channel 
decay into the state $a$ with $k$ electrons in the ballistic channel.
It takes place due to one-electron tunneling from the quantum dots 
into the ballistic channel with the rates $\Gamma_M$ and 
$\bar\Gamma_M$ respectively. This process is described by 
the second and the third terms in Eq.~(\ref{a13a}). Also  
the state $c$ (Fig. 4c) with $n-1$ electrons in the 
right reservoir can decay into the state $a$ due to tunneling 
to the right reservoir with the rate $\Gamma_R$ 
(the fourth term in Eq.~(\ref{a13a})).  The last term in this 
equation describes the decay of coherent superposition of the 
states $b$ and $c$ into the state $a$. It takes place due to single 
electron tunneling from the first and the second dots into the same state
of the ballistic channel with the amplitudes $\Omega_M$ and $\bar\Omega_M$,
respectively. Obviously, this process has no classical analogy,
since classical particle cannot simultaneously occupy two dots.
 
Equations (\ref{a13b}), (\ref{a13c}) and (\ref{a13d}) describe the 
probability rate of finding the system in the states
where one of the dots or both dots are occupied. In the first case
an electron can jump into unoccupied dot via continuum states of the 
ballistic channel. As a result, the states $b$ and $c$ can decay into 
linear superposition of the states $b$ and $c$. This process is 
described by the last terms in Eqs.~(\ref{a13b}), (\ref{a13c}).
Obviously, if the both dots are occupied, such a process cannot 
take a place. Therefore  
$\sigma_{dd}$ is not coupled with the nondiagonal density-matrix 
elements, Eq.~(\ref{a13d}). The last equation, (\ref{a13e}) describes 
the time-dependence of the nondiagonal density matrix element. 
It has the same interpretation as all previous equations. 

Equations (\ref{a13}) for the reduced density matrix $\sigma(t)$ 
were derived starting from the wave function $|\Psi (t)\rangle$, 
Eq.~(\ref{a2}), instead of using the density matrix for the entire 
system. Of course, it makes no difference if the
entire system is initially in the pure state, or all the levels of 
the reservoirs are occupied up to the corresponding Fermi energies. 
At finite temperature, however, the system is not initially in a pure 
state. Then one needs to perform the derivation using  
the Liouville (Landau--von Neumann) equation for 
the density-matrix. Yet, if energy levels $E_{1,2}$ of the 
dots are far away from the Fermi-levels
($T\ll E_F^L-E_{1,2},\; E_{1,2}-E_F^R$), then the reservoir levels 
that carry the current ($|E_{l}-E_{1,2}|\lesssim \Gamma$)
are deeply inside the Fermi sea. In this case one can consider these 
levels as fully occupied. Then neglecting the relaxation processes in the 
reservoir and assuming zero temperature inside 
the dots, we would arrive to the same Eqs.~(\ref{a13}), derived from 
the pure state. 

Using Eqs.~(\ref{a13}) one finds for the total current, 
Eq.~(\ref{a8})
\begin{equation}
I(t)=\Gamma_R[\sigma_{cc}(t)+\sigma_{dd}(t)],
\label{a14}
\end{equation}
where $\sigma_{ii}=\sum_{k,n}\sigma^{(k,n)}_{ii}$ are the total 
``probabilities''. We can easily understand this result
by taken into account that  $\sigma_{cc}+\sigma_{dd}$ is the 
total probability for occupation of the second dot and 
$\Gamma_R$ is the rate of electron transitions from this
dot to the (adjacent) right reservoir, Eq.~(\ref{c4}).

In order to find differential equations for $\sigma_{ij}$ we sum 
over $k,n$ in Eqs.~(\ref{a13}). Then we obtain the following 
Bloch-type equations, which describe the time-dependence of the 
density-matrix for separated dots 
\begin{mathletters}
\label{a15}
\begin{eqnarray}
&&\dot{\sigma}_{aa} = - \Gamma_L \sigma_{aa}
+\Gamma_M\sigma_{bb}+\bar\Gamma_M\sigma_{cc}
+\Gamma_R\sigma_{cc}
+2\pi\rho_M\Omega_M\bar\Omega_M(\sigma_{bc}+\sigma_{cb})
\label{a15a}\\
&&\dot{\sigma}_{bb} = - \Gamma_M \sigma_{bb}
+\Gamma_L\sigma_{aa}+\bar\Gamma_M\sigma_{dd}
+\Gamma_R\sigma_{dd}
-\pi\rho_M\Omega_M\bar\Omega_M(\sigma_{bc}+\sigma_{cb})
\label{a15b}\\
&&\dot{\sigma}_{cc} = -(\Gamma_L +\bar\Gamma_M +\Gamma_R) 
\sigma_{cc}
+\Gamma_M\sigma_{dd}
-\pi\rho_M\Omega_M\bar\Omega_M(\sigma_{bc}+\sigma_{cb})
\label{a15c}\\
&&\dot{\sigma}_{dd} = -(\Gamma_M +\bar\Gamma_M +\Gamma_R) 
\sigma_{dd}+\Gamma_L\sigma_{cc}
\label{a15d}\\
&&\dot{\sigma}_{bc} = i(E_2-E_1)\sigma_{bc}
-\pi\rho_M\Omega_M\bar\Omega_M(\sigma_{bb}+\sigma_{cc})
+2\pi\rho_M\Omega_M\bar\Omega_M\sigma_{dd}-\frac{1}{2}
\Gamma_{tot}\sigma_{bc},
\label{a15e}
\end{eqnarray}
\end{mathletters}
where $\Gamma_{tot}=\Gamma_L+\Gamma_M +\bar\Gamma_M +\Gamma_R$.

The stationary (dc) current $I=I(t\to\infty )$ 
Eq.~(\ref{a14}) can be easily obtained from Eqs.~(\ref{a15}) by taken 
into account that $\dot\sigma_{ij}\to 0$ for $t\to\infty$. Then 
Eqs.~(\ref{a15}) turn into a system of linear algebraic equations, 
supplemented by a probability conservation condition 
$\sigma_{aa}+\sigma_{bb}+\sigma_{cc}+\sigma_{dd}=1$. Consider, for example,
the case of the same partial widths of the levels $E_{1,2}$, i.e. 
$\Gamma_L=\Gamma_R=\Gamma_0$ and $\Gamma_M=\bar\Gamma_M$. The latter 
implies $2\pi\rho_M\Omega_M\bar\Omega_M=\pm \Gamma_M$, since the amplitudes 
$\Omega_M$, $\bar\Omega_M$ can be of the opposite signs. Solving 
Eqs.~(\ref{a14}), (\ref{a15}) one finds for dc current
\begin{equation}
I=\frac{\Gamma_0^2\Gamma_M^2}{2\epsilon^2(\Gamma_0+\Gamma_M)
+2\Gamma_0(\Gamma_0^2+3\Gamma_0\Gamma_M+3\Gamma_M^2)},
\label{a16}
\end{equation} 
where $\epsilon =E_1-E_2$.
 
Similar to the coupled-dot case, Eq.~(\ref{c2}), dc current 
in separated dots displays the Lorentzian shape resonance as 
a function of $\epsilon$ and the same peculiar dependence on the 
coupling with the leads. The latter manifests the quantum-coherence 
effects in separated dot systems. We discuss these effects
in the next section. 

\section{Coherent effects in two-dot systems}
It follows from the rate equations (\ref{a15}) 
that the diagonal density-matrix elements
are coupled with the non-diagonal terms  ($\sigma_{bc}$, $\sigma_{cb}$),
similar to Eqs.~(\ref{c1}) for a coupled-dot system. It means that the  
coherent effects should survive even in separated quantum dots. 
Indeed, let us consider the dependence of the resonant current on 
a coupling with the reservoirs.  
As an example, we show in Fig. 5a the peak value of the resonant current
($\epsilon =0$) obtained from Eqs.~(\ref{a14}), (\ref{a15}) 
as a function of $\Gamma_R$ at fixed $\Gamma_L=\Gamma_M$. 
One finds that the current through the dots separated 
by a ballistic channel displays    
the same peculiar behavior with $\Gamma_R$, as in the case 
of coupled dots, Fig. 3. Namely, the current {\em decreases} 
when the penetrability of the right barrier {\em increases}. 
Such a behavior of the resonant 
current is a clear manifestation of the quantum coherence 
effects, as explained in Sect. II (see also\cite{gm,gbj,bg}).
It also can be verified experimentally by changing the voltage
on the gate, connecting the right dot with the collector.
 
On the other hand, the dependence of the resonant current on the width 
$\Gamma_M$ is quite different, Fig. 5b. One finds that the current 
increases with $\Gamma_M$. It is not surprising, since $\Gamma_M$ 
plays a role of coupling between the two dots. In fact, a similar
behavior with the coupling $\Omega_0$ displays the resonant current 
in the case of coupled dots, Eq.~(\ref{c2}). 

Despite the analogues quantum coherence effects, 
the rate equations (\ref{a15}) are different from their counterpart   
for the coupled-dots, Eqs.~(\ref{c1}). One finds that the diagonal 
matrix elements in Eqs.~(\ref{a15}) are coupled with 
the {\em real} part of the non-diagonal
matrix element, 
\vskip1cm 
\begin{minipage}{13cm}
\begin{center}
\leavevmode
\epsfxsize=13cm
\epsffile{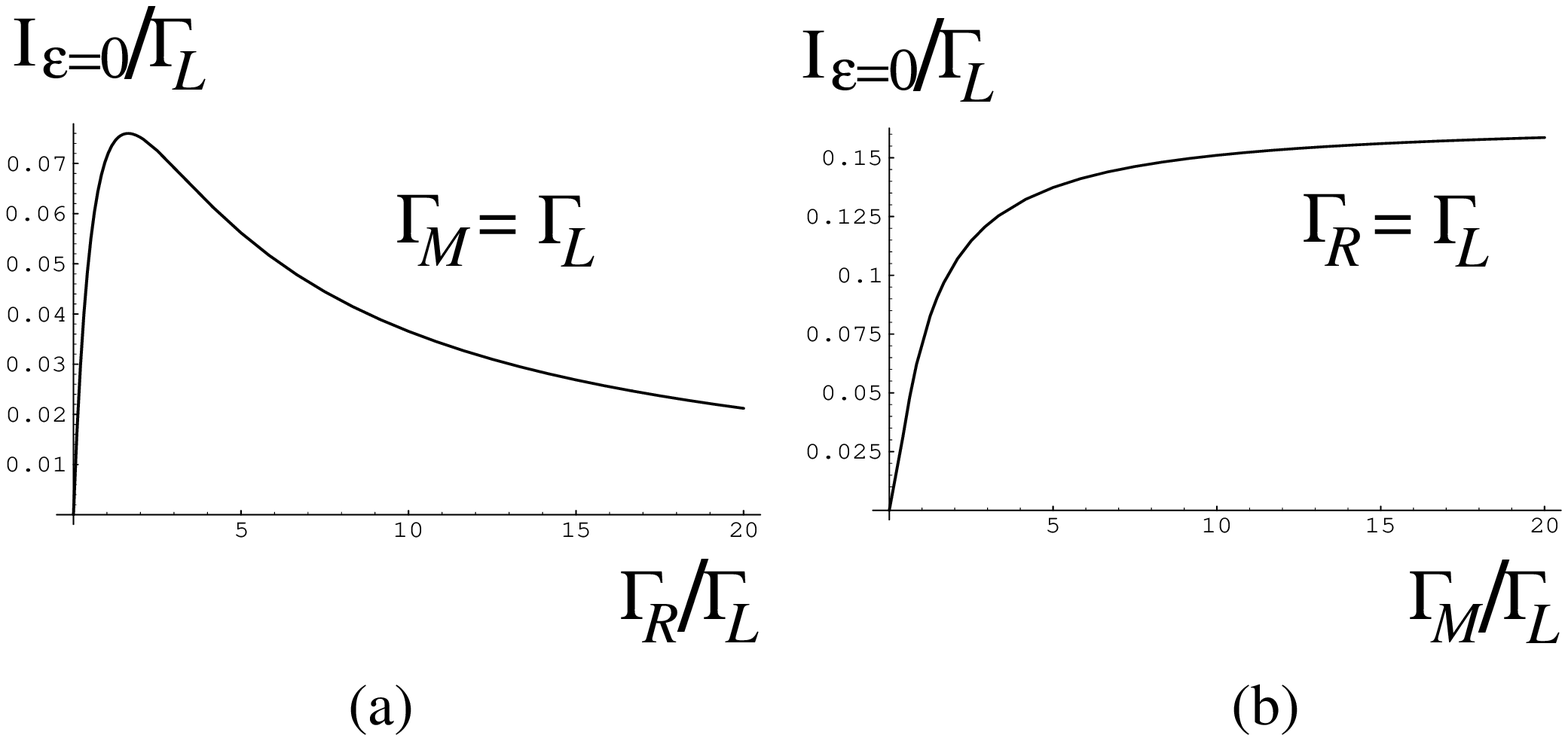}
\end{center}
{\begin{small}
Fig.~5. DC current through the separated dot structure, Fig. 2, as
a function of $\Gamma_R/\Gamma_L$ for $\Gamma_M=\Gamma_L$ (a), and 
as a function of $\Gamma_M/\Gamma_L$ for $\Gamma_R=\Gamma_L$ (b).
\end{small}}
\end{minipage} \\ \\ 
whereas the corresponding terms in Eqs.~(\ref{c1}),
are  coupled with the {\em imaginary} part
of the non-diagonal matrix element. Notice also that the coupling 
is determined by the hopping amplitude 
$\Omega$ in Eqs.~(\ref{c1}), while the corresponding coupling 
in Eqs.~(\ref{a15}) is proportional to $\Omega^2\rho$ 

It is important to point out that the coupling with the 
non-diagonal density-matrix elements in Eqs.~(\ref{a15}) 
does not decrease with the separation distance between the dots,
although the hopping amplitudes $\Omega_M,\,\bar\Omega_M$ 
do decrease. Indeed, using semi-classical expressions one finds
$|\Omega_M|=(1/\sqrt{\tau_1\tau_M})\exp (-S)$ and 
$|\bar\Omega_M|=(1/\sqrt{\tau_2\tau_M})\exp (-\bar S)$, where
$S(\bar S)$ is the action under the barrier, separating the first  
(second) dot from the ballistic channel. It can be written as  
$S=\int_{x_i}^{x_f}|p(x)|dx$, where $|p(x)|=\sqrt{2m[V(x)-E]}$ 
and $x_{i,f}$, are the classical turning points. 
$\tau_{1,2}$ are the classical periods of motion 
in the first (second) quantum well 
and $\tau_M$ is the classical periods of motion 
in the ballistic channel. It implies that $\tau_M\to\infty$ when the length 
of the ballistic channel increases.  Yet, $\tau_M$ is canceled 
out from the coupling between the diagonal and non-diagonal matrix elements,
since the ballistic channel is effectively 
one-dimensional so that the density of states  
is $\rho_M=\tau_M/2\pi$. Finally we obtain 
\begin{equation}
2\pi\rho_M\Omega_M\bar\Omega_M=\pm 
\frac{1}{2\pi\sqrt{\tau_1\tau_2}}\exp [-(S+\bar S)].
\label{b1}
\end{equation} 
Therefore, the quantum coherence effects in dc current could 
survive even for long ballistic channels. 
Obviously, it is true only for ideal one-dimensional channels, where 
all the flux coming from the first dot arrives the second dot.
It should be also pointed out that electron-electron 
scattering inside the ballistic channel (not taken into account in 
these calculations) may reduce the interference effects. Yet, we 
expect that such a process would not diminish the coupling 
$(\ref{b1})$, but increase the ``dissipation'' 
width $\Gamma_{tot}$ in Eq.~(\ref{a15e}).  
This problem, however, needs a special investigation. 

It is interesting to make a comparison between peak values 
of the dc current in coupled and separated dots. 
In both cases the current reaches its maximal value 
for $\Gamma_L=\Gamma_R=\Gamma_0$ and $\epsilon=0$, Eqs.~(\ref{c2}), 
(\ref{a16}).  One finds  for the coupled dots  
\begin{equation}
I_{max}=\frac{\Gamma_0}{2}
\frac{\Omega_0^2}{\Omega_0^2+\Gamma_0^2/4}\rightarrow
\frac{\Gamma_0}{2},\quad\quad {\mbox{for}}\quad\quad \Omega_0\gg \Gamma_0,  
\label{b2}
\end{equation}
while for the separated dots 
\begin{equation}
I_{max}=\frac{\Gamma_0}{2}
\frac{\Gamma_M^2}{3\Gamma_M^2+3\Gamma_0\Gamma_M
+\Gamma_0^2}\rightarrow
\frac{\Gamma_0}{6},\quad\quad {\mbox{for}}\quad\quad \Gamma_M\gg \Gamma_0,  
\label{b3}
\end{equation}
Thus the peak value of dc current 
in the coupled dots is three times larger than that 
in the separated dots. 

It is quite natural to associate quantum coherence effects in double-well 
systems with quantum oscillations of an electron between the wells. 
Indeed, an electron in a coupled-dot system, detached from the 
emitter and the collector, oscillates between the states (a) and (b), 
Fig. 6. These oscillations are reproduced by the same rate equations 
that describe quantum transport in coupled-dots, Eqs.~(\ref{c1}),
but now with $\Gamma_L=\Gamma_R=0$. We then obtain for 
the probability of finding an electron in the states (a) and (b), 
$\sigma_{aa}(t)$ and $\sigma_{bb}(t)$,  Fig.6:
\begin{mathletters}
\label{b33}
\begin{eqnarray}
&&\dot\sigma_{aa}=i\Omega_0(\sigma_{ab}-\sigma_{ba})
\label{b33a}\\
&&\dot\sigma_{bb}=-i\Omega_0(\sigma_{ab}-\sigma_{ba})
\label{b33b}\\
&&\dot\sigma_{ab}=i(E_2-E_1)\sigma_{ab}
+i\Omega_0(\sigma_{aa}-\sigma_{bb}).
\label{b33c}
\end{eqnarray}
\end {mathletters}
\vskip1cm 
\begin{minipage}{13cm}
\begin{center}
\leavevmode
\epsfxsize=6cm
\epsffile{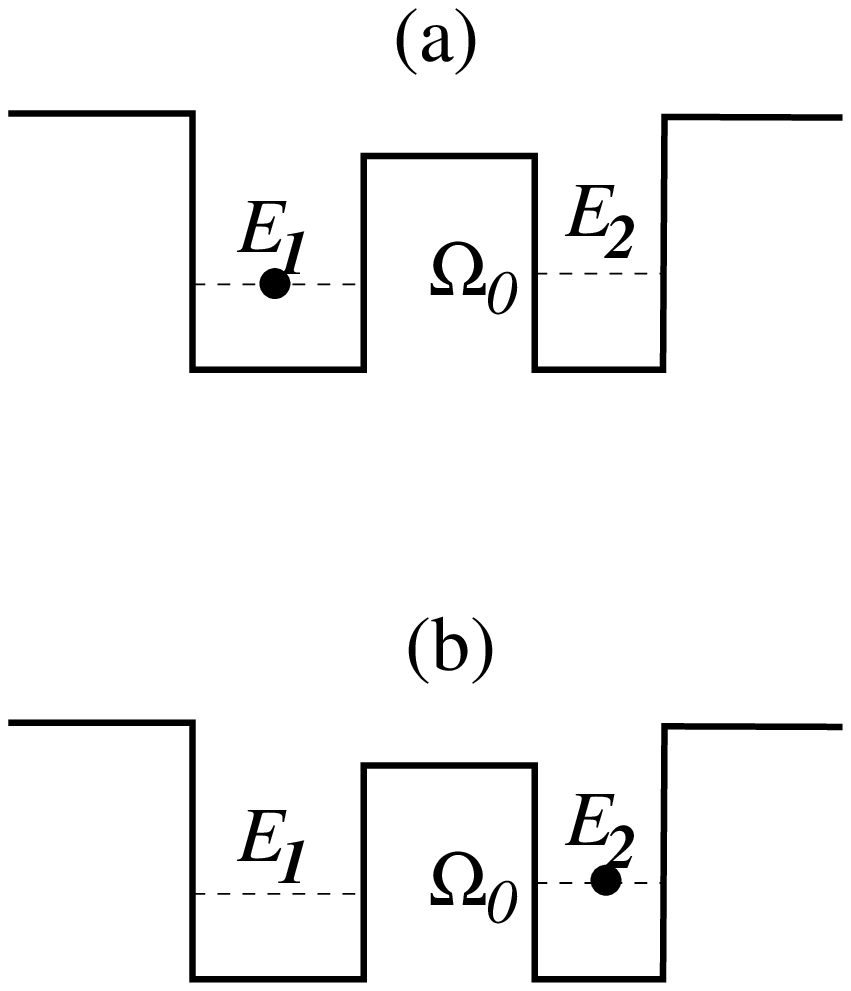}
\end{center}
{\begin{small}
Fig.~6. Electron states of the coupled-dot structure of Fig. 1, 
isolated from the leads.
\end{small}}
\end{minipage} \\ \\ 
Solving this equations with the initial condition 
$\sigma_{aa}(0)=1$ and $\sigma_{bb}(0)=0$, we obtain
\begin{equation}
\sigma_{aa}(t)=\frac{\Omega_0^2\cos^2(\omega_1t)+\epsilon^2/4}
{\Omega_0^2+\epsilon^2/4}
\label{b4}
\end{equation}
where $\omega_1=(1/2)\sqrt{4\Omega_0^2+\epsilon^2}$. 
As expected, an electron 
oscillates between the dots with an amplitude $1/(1+\epsilon^2/4\Omega_0^2)$.

The situation is different, however, if the dots are separated 
by a ballistic channel, Fig.~7. Indeed, the corresponding rate 
equations for the probabilities 
$\sigma_{aa}(t)$ and $\sigma_{bb}(t)$ 
of finding an electron in the states (a) and (b) of Fig.~7 are obtained from 
Eqs.~(\ref{a15}) for $\Gamma_L=\Gamma_R=0$. Consider 
for simplicity the case of $\Omega_M=\bar\Omega_M$. Then 
Eqs.~(\ref{a15}) become 
\begin{mathletters}
\label{b44}
\begin{eqnarray}
&&\dot\sigma_{aa}=-\Gamma_M\sigma_{aa}-\frac{1}{2}\Gamma_M
(\sigma_{ab}+\sigma_{ba})
\label{b44a}\\
&&\dot\sigma_{bb}=-\Gamma_M\sigma_{bb}-\frac{1}{2}\Gamma_M
(\sigma_{ab}+\sigma_{ba})
\label{b44b}\\
&&\dot\sigma_{ab}=i(E_2-E_1)\sigma_{ab}-\frac{1}{2}\Gamma_M
(\sigma_{aa}+\sigma_{bb})-\Gamma_M\sigma_{ab}
\label{b44c}
\end{eqnarray}
\end {mathletters}
Solving  these equations with 
the initial condition $\sigma_{aa}(0)=1$ and $\sigma_{bb}(0)=0$, we obtain
\begin{equation}
\sigma_{aa}(t)=\frac{\Gamma_M^2\cosh^2(\omega_2t)
-\epsilon^2}{\Gamma_M^2-\epsilon^2}\exp (-\Gamma_Mt),
\label{b5}
\end{equation}
where $\omega_2=(1/2)\sqrt{\Gamma_M^2-\epsilon^2}$.
It follows from Eq~(\ref{b5}) that $\sigma_{aa}(t)$ does not display any  
\vskip1cm 
\begin{minipage}{13cm}
\begin{center}
\leavevmode
\epsfxsize=12cm
\epsffile{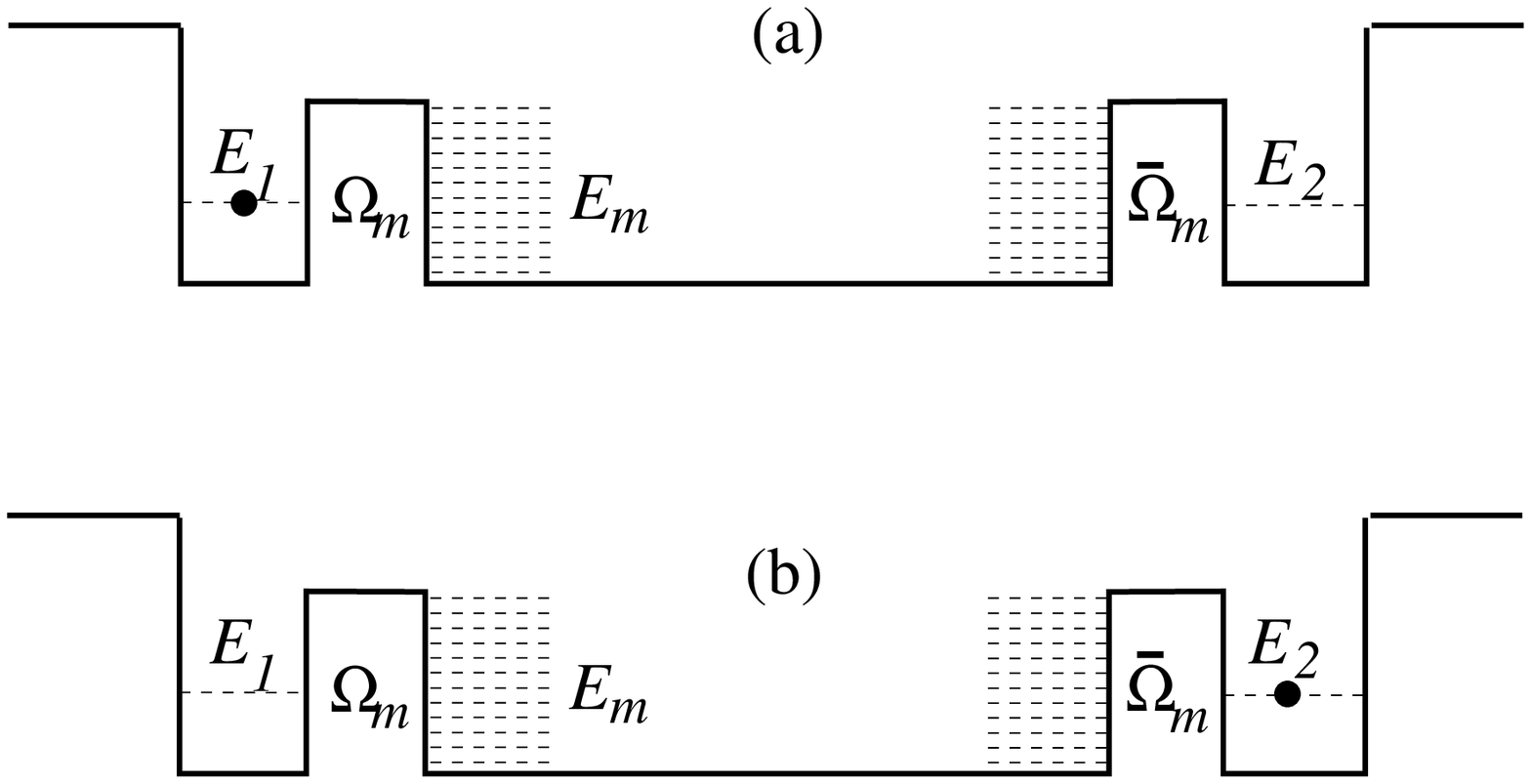}
\end{center}
{\begin{small}
Fig.~7. Electron states of the two separate dot structure of Fig. 2, 
isolated from the leads.
\end{small}}
\end{minipage} \\ \\ 
oscillations. As an example, we show in Fig. 8 the probability 
of finding an electron in   
the left dot as a function of time in the case of coupled and separated 
dots (Figs.~6, 7) for  $\epsilon =\Gamma_M =\Omega_0$.
The solid line corresponds to Eq.~(\ref{b4}), and 
the dashed line to Eq.~(\ref{b5}).
Notice that $\sigma_{aa}=\sigma_{bb}\to 1/4$
for $t\to\infty$ when the levels are aligned ($\epsilon=0$), and
$\sigma_{aa}=\sigma_{bb}\to 0$ for $t\to\infty$ when $\epsilon\not = 0$.
It means that an electron decays into the ballistic channel.

\vskip1cm 
\begin{minipage}{13cm}
\begin{center}
\leavevmode
\epsfxsize=7cm
\epsffile{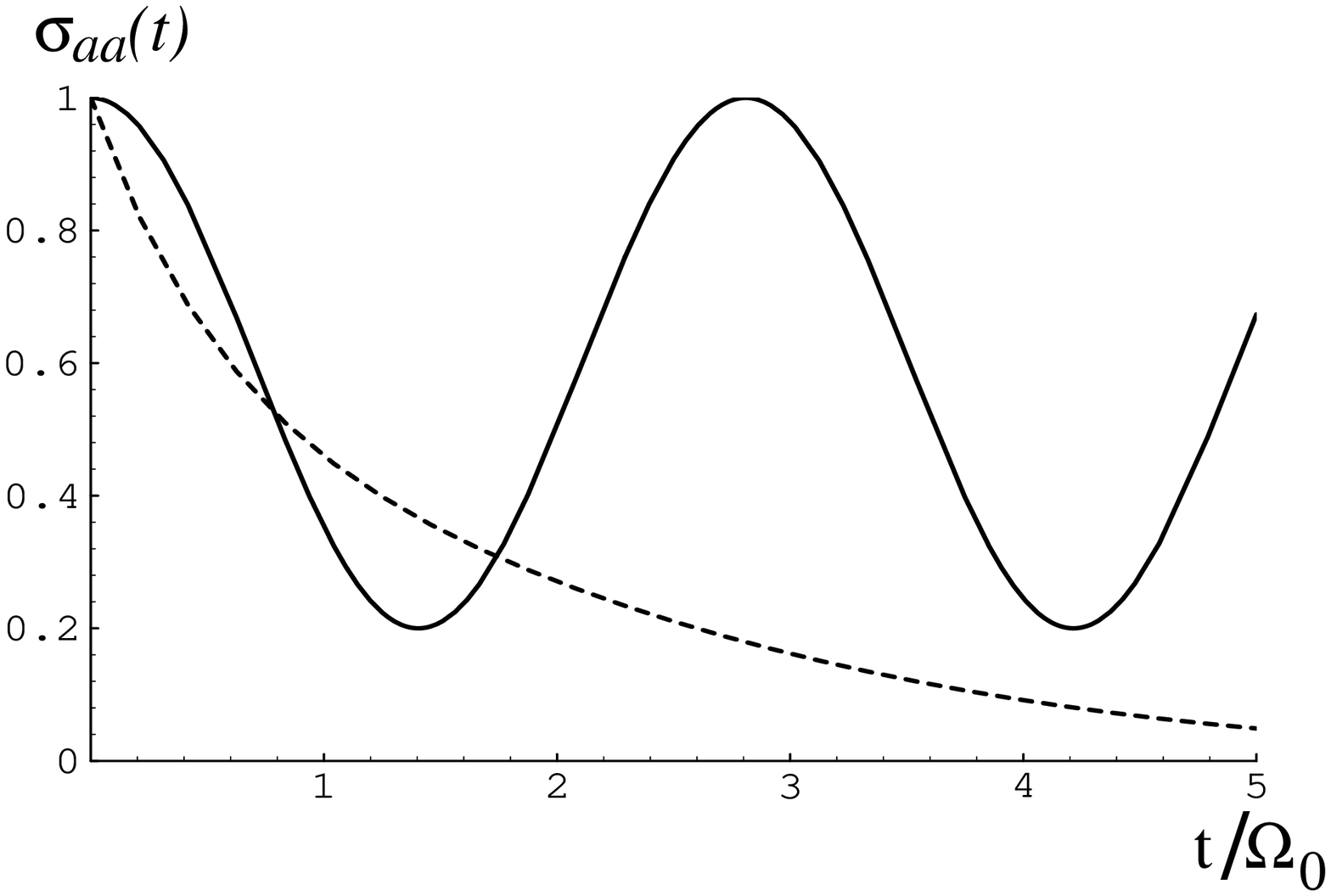}
\end{center}
{\begin{small}
Fig.~8. The occupation of the left dot in Figs. 6 and 7  
as a function of time for $\epsilon =\Gamma_M =\Omega_0$. The 
solid line corresponds to coupled dots, and the 
dashed line to separated dots. 
\end{small}}
\end{minipage} \\ \\ 

The reason for a dissipative behavior of $\sigma_{aa}$ 
in the case separated dots is a coupling of 
the dots with continuum states of the ballistic 
channel. The latter leads to dephasing that destroys 
quantum oscillations. It is thus rather remarkable that the
quantum coherence effects would ``reappear'' when 
the same system is connected with the leads, Fig.~2.

\section{General case} 
The rate equations (\ref{a15}), describing electron  
transport in separated dots can be extended to any multi-dot system. 
By applying the same technique of integrating out the reservoir 
states as in Ref.\cite{gp} and in Appendix, we   
arrive to the rate equations for the density-matrix $\sigma_{\alpha\beta}$ 
of the multi-dot system. These equations can be written as 
\begin{eqnarray}
\dot\sigma_{\alpha\beta} &=&  i(E_\beta - E_\alpha) \sigma_{\alpha\beta} +
i\left (\sum_{\gamma}\sigma_{\alpha\gamma}
\tilde\Omega_{\gamma\to\beta}
-\sum_{\gamma}\tilde\Omega_{\alpha\to\gamma}
\sigma_{\gamma\beta}\right )\nonumber\\
&-&{\sum_{\gamma,\delta}}^{\prime}
\pi\rho(\sigma_{\alpha\gamma}\Omega_{\gamma\to\delta}\Omega_{\delta\to\beta}
+\sigma_{\gamma\beta}\Omega_{\gamma\to\delta}\Omega_{\delta\to\alpha})
+\sum_{\gamma,\delta}
\pi\rho\,(\Omega_{\gamma\to\alpha}\Omega_{\delta\to\beta}+
\Omega_{\gamma\to\beta}\Omega_{\delta\to\alpha})
\sigma_{\gamma\delta}\, .
\label{d1}
\end{eqnarray}
Here $|\alpha\rangle,\,  |\beta\rangle,\ldots$ are the states of 
the multi-dot system in the occupation number representation, and   
$\Omega_{\alpha\to\beta}$ denotes one-electron 
hopping amplitude that generates $\alpha\to\beta$-transition. We distinguish 
between the amplitudes $\tilde\Omega$ and $\Omega$ of   
one-electron hopping among isolated states and among isolated
and continuum states, respectively. The latter transitions are 
of the second order in the hopping amplitude $\sim\Omega^2$. 
These transition are produced by two consecutive hoppings of an electron with 
the same energy across continuum states with the density of states $\rho$.      
The symbol $\sum'$ means that only those transitions 
$\gamma\to\delta\to (\alpha ,\beta)$ are accounted for
in the sum, where the number of electrons in continuum 
states (in the reservoirs and ballistic channels) stays the 
same in the initial and final states (c.f. Eqs.~(\ref{a13})).  

It is rather easy to verify that Eqs.~(\ref{d1}) coincide with 
Eqs.~({\ref{a15}) for $\alpha,\beta,\ldots = \{a,b,c,d\}$,
which are the states of the separated dot system, shown in Fig.~5.  
Let us compare now Eqs.~(\ref{d1}) with the Bloch rate 
equations (\ref{c3}) for quantum 
transport in coupled dots. We begin with the  
equations for diagonal density-matrix elements, $\beta=\alpha$. 
The first term in Eq.~(\ref{d1}) is zero, since $E_\alpha=E_\beta$.
The first term in  Eq.~(\ref{c3a}) and the second term in 
Eq.~(\ref{d1}) have the same 
form and describe the coupling with non-diagonal matrix elements 
generated by one-electron hopping between isolated states. 
The remaining terms in both equations look differently. 
However, one can easily realize that in the case of coupled dots the only
possible transitions for diagonal matrix elements 
are those corresponding to $\gamma =\alpha$. 
Then the third term of Eq.~ (\ref{d1}) becomes 
\begin{equation}
\sum_{\delta}
2\pi\rho\Omega_{\alpha\to\delta}\Omega_{\delta\to\alpha}
\sigma_{\alpha\alpha}=
\sigma_{\alpha\alpha}\sum_{\delta}\Gamma_{\alpha\to\delta}.
\label{d2}
\end{equation}
It coincides  with the second (``dissipative'') term 
of the Bloch rate equation (\ref{c3a}).
The last term in Eq.~(\ref{d1}) describes the reverse process, i.e.  
all possible conversions of the states $\gamma$,  $\delta$
into the state $\alpha$. Again, this term turns into the third term of
Eq.~(\ref{c3a}) for   $\gamma =\delta$. 

The same relation can be traced for the non-diagonal 
density-matrix elements. We find that the third term of Eq.~(\ref{d1}), which  
describe all possible decays of the states $\alpha$ 
and $\beta$, reproduce the third term in Eq.~(\ref{c3b}) 
for $\gamma=\beta$ and $\delta=\alpha$, respectively. The fourth term 
in Eq.~(\ref{d1}) describes the reverse process, i.e. 
the conversion of $\sigma_{\gamma\delta}\to\sigma_{\alpha\beta}$
produced by two consecutive hoppings through continuum state 
media.  In particular, if $\Omega_{\gamma\to\alpha}=\Omega_{\delta\to\beta}$,
these terms reproduce the last term of Eq.~(\ref{c3b}), where  
$\Gamma_{\gamma\delta\rightarrow \alpha\beta}=
2\pi\rho\,\Omega_{\gamma\to\alpha}\Omega_{\delta\to\beta}$. 

\section{summary}
In this paper we derived quantum rate equations, which provide 
the most simple and transparent way for a description of the both 
coherent and incoherent electron transport in quantum dot systems.
In the beginning we considered    
a system of two quantum dots linked by a ballistic channel and 
connected with the emitter and collector reservoirs. 
Starting with the many-particle wave function 
and integrating out the continuum states,
we have obtained the equations of motion for the density
submatrix of the two-dot system in the occupation number representation. 
In spite of the quantum dots might be away one from another,
we found that the diagonal density-matrix elements are still 
coupled with the non-diagonal density-matrix elements, similar 
to Bloch equations for double-well  systems. 
The essential difference, however, is that 
the diagonal matrix elements in the Bloch equations are 
coupled with the {\em imaginary} part 
of the non-diagonal density-matrix elements, while 
for the separated wells the corresponding terms  
are coupled with the {\em real} part of the non-diagonal 
density-matrix elements. 

As expected, the coupling between diagonal and 
non-diagonal matrix elements generated quantum coherent effects
in electron transport. For instance, we found that due to 
destructive quantum interference the dc current in separated dots
is reduced when the barrier penetrability increases. 
This effect would survive even if the dots are largely separated,   
providing that they are linked 
by ideal one-dimensional ballistic channels. 

In spite of the quantum coherence effects found in dc current, the same 
system of separated dot, detached from the emitter and 
collector reservoirs does not show any quantum oscillations. 
It is quite different from a 
coupled-dot system with aligned levels, 
in which an electron oscillates between the dots. 
 
The rate equations for two separated quantum dots
are extended to a general case of multi-dot system, 
in which the dots are either directly coupled, or interconnected 
via ballistic channels.
These new rate equations generalize the 
well-known Bloch equations, describing time-evolution of the 
density-matrix of coupled 
multi-well systems in the presence of a dissipative media. 
We thus expect that the applicability of 
our generalized rate equations is not restricted by    
quantum transport in multi-dot system only, 
but these equations would be very useful for 
various physical problems where quantum coherence and classical 
dissipation effects do interplay.  
\section{Acknowledgments}
I thank Yu. Nazarov for valuable discussions. 

\appendix
\section{Derivation of rate-equations for separated dots}
We demonstrate of how the sum over continuum states in  
the density matrix, Eqs.~(\ref{a7}).
can be performed analytically leading to the rate equations 
for two-dot systems.  
In order to simplify the algebraic transformations  
we use the Laplace transform for the Shr\"odinger equation 
$i|\dot\Psi (t)\rangle ={\cal H}|\Psi (t)\rangle$. Then the 
amplitudes $b(t)$ in the wave function (\ref{a2}) are replaced  
by their Laplace transform
\begin{equation}
\tilde{b}(E)=\int_0^{\infty}e^{iEt}b(t)dt\, ,
\label{a3}
\end{equation}
Substituting Eq.~(\ref{a2}) into the Shr\"odinger equation 
we obtain an infinite set of coupled equations for the 
amplitudes $\tilde b(E)$:
\begin{mathletters}
\label{a4}
\begin{eqnarray}
& &E \tilde{b}_{0}(E) - \sum_l \Omega_{l}\tilde{b}_{1l}(E)=i
\label{a4a}\\
&(&E + E_{l} - E_1) \tilde{b}_{1l}(E) - \Omega_{l}
      \tilde{b}_0(E) -\sum_m\Omega_{m}\tilde{b}_{lm}(E)=0
\label{a4b}\\
&(&E + E_{l} - E_m) \tilde{b}_{lm}(E)-\Omega_m\tilde{b}_{1l}(E)
-\bar\Omega_m\tilde{b}_{2l}(E)-\sum_{l'}\Omega_{l'}\tilde{b}_{12ll'}(E)=0
\label{a4c}\\
&(&E + E_{l} - E_2) \tilde{b}_{2l}(E) -
      \sum_m\bar\Omega_m \tilde{b}_{lm}(E) - 
      \sum_{r} \Omega_{r}\tilde{b}_{lr}(E)-
      \sum_{l'} \Omega_{l'}\tilde{b}_{12ll'}(E)=0
\label{a4d}\\
&(&E + E_{l} + E_{l'} - E_1 - E_2-U_{12}) \tilde{b}_{12ll'}(E)-
\Omega_{l'} \tilde{b}_{2l}(E)+
\Omega_{l} \tilde{b}_{2l'}(E)-\nonumber\\
&&~~~~~~~~~~~~~~~~~~~~~~~~~~
-\sum_{m} \bar\Omega_{m}\tilde{b}_{1ll'm}(E)
-\sum_{m} \Omega_{m}\tilde{b}_{2ll'm}(E)-
\sum_{r} \Omega_{r}\tilde{b}_{1ll'r}(E)=0
\label{a4e}\\
& &\cdots\cdots\cdots\cdots\cdots 
\nonumber
\end{eqnarray}
\end{mathletters}
Notice that due to the Pauli principle an electron can return back  
only into unoccupied states of the emitter. As a result, the summation  
over the emitter states does not appear in the corresponding terms
of Eqs.~(\ref{a4}) (in the second term of Eq.~(\ref{a4b}), 
in the second and the third terms of Eq.~(\ref{a4c}), and so on).     

Eqs. (\ref{a4}) can be substantially simplified. Let us replace  
the amplitude $\tilde b$ in the term $\sum\Omega\tilde b$ 
of each of the equations  by 
its expression obtained from the subsequent equation. For example,   
substitute $\tilde{b}_{1l}(E)$ from Eq.~(\ref{a4b}) into Eq.~(\ref{a4a}).
We obtain
\begin{equation}
\left [ E - \sum_l \frac{\Omega^2_L(E_l)}{E + E_{l} - E_1}
    \right ] \tilde{b}_{0}(E) - \sum_{l,m}
    \frac{\Omega_L(E_l)\Omega_M(E_m)}{E + E_{l} - E_1}
    \tilde{b}_{lm}(E)=i,
\label{a5}
\end{equation}
where $\Omega_l\equiv \Omega_L(E_l)$ and 
$\Omega_m\equiv \Omega_M(E_m)$. 
Since the states in the reservoirs are very dense (continuum), 
one can replace the sums over $l$ and $m$ by integrals, for instance  
$\sum_{l}\;\rightarrow\;\int \rho_{L}(E_{l})\,dE_{l}\:$,
where $\rho_{L}(E_{l})$ is the density of states in the emitter. 
Then the first sum in Eq.~(\ref{a5}) becomes an
integral which can be split into a sum of the singular and principal value 
parts. The singular part
yields $\;\,-i\Theta (E_F^L+E-E_1)\,\Gamma_L/2$, where $\Gamma_L = 2\pi
\rho_L(E_1)|\Omega_L(E_1)|^2$ is the level $E_1$ partial width 
due to coupling to the emitter. Let us assume that  
$E_F^L\gg E_1\gg E_F^M$, i.e.
the energy level is deeply inside the band.  
In this case the integration over $E_{l(m)}$-variables can be extended 
to $\pm\infty$. As a result, the theta-function can be replaced by one, and   
the principal part is merely included into
redefinition of the energy $E_1$.  
The second sum (integral) in Eq.~(\ref{a5}) proves to be
negligible small. Indeed, let us replace   
$\tilde{b}_{lm}\to\tilde{b}(E_l,E_m,E)$, and assume weak energy
dependence of $\Omega$ on $E_{l}$. Then 
one finds from Eqs. (\ref{a4}) that the poles of the integrand in the 
$E_l$-variable are on one side of the integration contour, 
and therefore this term vanishes. In general, any terms of 
the type $\int\cdots dE_s\cdots \tilde b(\cdots ,E_s,\cdots )
(E+\cdots\pm E_s)^{-1}\to 0$, whenever the integration over 
the $E_s$-variable can be extended to $\pm \infty$. We shall imply
this property in all subsequent derivations. Notice 
that these results are valid also for non-zero temperature, providing 
that $T\ll E_F^L-E_{1,2},\; E_{1,2}-E_F^M$. 

Now we apply analogous considerations to the other equations of the
system (\ref{a4}). It follows from Eqs.~(\ref{a4}) that the  
Coulomb repulsion just shifts the energy levels in Eq.~(\ref{a5}),
$E_{1,2}\to E_{1,2}+U_{ij}$. If the shifted 
level is outside the band, the singular part of the integral vanishes 
and therefore the corresponding width $\Gamma$
is replaced by zero. If, however, the shifted energy stands 
deeply inside the band, all the previous treatment remains 
the same. A problem appears only if the shifted energy 
in near the Fermi level. This case is not considered here.   
Finally we arrive at the following set of equations: 
\begin{mathletters}
\label{a6}
\begin{eqnarray}
&& (E + i\Gamma_L/2) \tilde{b}_{0}=i
\label{a6a}\\
&& (E + E_{l} - E_1 + i\Gamma_M/2) \tilde{b}_{1l}
      - \Omega_{l} \tilde{b}_{0}
+i\pi\rho_M\Omega_M\bar\Omega_M\tilde{b}_{2l}=0
\label{a6b}\\ 
&& (E + E_{l} - E_{m} + i\Gamma_L/2) \tilde{b}_{lm} -
      \Omega_{m} \tilde{b}_{1l}-\bar\Omega_{m} \tilde{b}_{2l}=0
\label{a6c}\\
&& (E + E_{l} -E_2+ i\Gamma'_L/2+ i\bar\Gamma_M/2+ i\Gamma_R/2) 
       \tilde{b}_{2l}+i\pi\rho_M\Omega_M\bar\Omega_M\tilde{b}_{1l}=0 
      \label{a6d}\\
&& (E + E_{l} + E_{l'} - E_1 - E_2 -U_{12}+
i\Gamma'_M/2+ i\bar\Gamma'_M/2+ i\Gamma'_R/2) \tilde{b}_{12ll'}-
\Omega_{l'} \tilde{b}_{2l}+
\Omega_{l} \tilde{b}_{2l'}=0,
\label{a6e}\\
& &\cdots\cdots\cdots\cdots\cdots 
\nonumber
\end{eqnarray}
\end{mathletters}
where $\Gamma_M = 2\pi\rho_M(E_1)|\Omega_M(E_1)|^2$ and 
$\bar\Gamma_M = 2\pi\rho_M(E_2)|\bar\Omega_M(E_2)|^2$ are 
the partial widths of the levels $E_1$ and $E_2$ respectively due 
to coupling to the ballistic channel and $\rho_M$ is the density 
of states in the ballistic channel.  
$\Gamma_R = 2\pi\rho_R(E_2)|\Omega_R(E_2)|^2$ is 
the level $E_2$ partial width due to coupling to the collector.
The Coulomb interdot repulsion $U_{12}$ just modifies the 
corresponding width ($\Gamma$ is replaced by $\Gamma'$, evaluated 
at the energy $E_{1,2}+U_{12}$) whenever both dots are 
occupied\cite{gp}. A similar modification would take place 
if we include the Coulomb repulsion between electrons inside 
the dots and the ballistic channel. In the following we 
assume that widths are weakly dependent on the energy,
so $\Gamma' =\Gamma$ everywhere.

The density matrix elements, Eqs.~(\ref{a7}), are directly related 
to the amplitudes $\tilde b(E)$ through the inverse Laplace transform 
\begin{equation}
\sigma^{(k,n)}(t)=
\sum_{l\ldots ,m,\ldots, r\ldots}
\int \frac{dEdE'}{4\pi^2}\tilde b_{l\cdots m\cdots r\cdots}(E)
\tilde b^*_{l\cdots m\cdots r\cdots}(E')e^{i(E'-E)t}
\label{a9}
\end{equation}
Using this equation one can transform  Eqs. (\ref{a6}) for 
the amplitudes $\tilde b(E)$  
into differential equations for the probabilities $\sigma^{(k,n)}(t)$.
Consider, for instance, the term 
$\sigma_{bb}^{(0,0)}(t)=\sum_l |b_{1l}(t)|^2$,
Eq.~(\ref{a7b}). Multiplying Eq.~(\ref{a6b}) by $\tilde{b}^*_{1l}(E')$ 
and then subtracting  the complex conjugated 
equation with the interchange 
$E\leftrightarrow E'$ we obtain 
\begin{eqnarray} 
&&\int\frac{dEdE'}{4\pi^2}\sum_l\left\{(E'-E
-i\Gamma_M)\tilde b_{1l}(E)\tilde b^*_{1l}(E')
-\Omega_l[\tilde b_{1l}(E)\tilde b^*_0(E')-\tilde b^*_{1l}(E')\tilde b_0(E)]
\right.\nonumber\\
&&\left.~~~~~~~~~~~~~~~~~~~~~~~~~~~~~~~~~~-i\pi\rho_M\Omega_M\bar\Omega_M
[\tilde b_{1l}(E)\tilde b^*_{2l}(E')+\tilde b^*_{1l}(E')\tilde b_{2l}(E)]
\right\}e^{i(E'-E)t}=0.
\label{a10}
\end{eqnarray}
Substituting 
\begin{equation}
\tilde b_{1l}(E)=\frac{\Omega_l\tilde b_0(E)-
i\pi\rho_M\Omega_M\bar\Omega_M\tilde b_{2l}(E)}
{E+E_l-E_1+i\Gamma_M/2}
\label{a11}
\end{equation}
into Eq.~(\ref{a10}) we can carry out the $E,E'$-integrations thus 
obtaining
\begin{equation}
\dot{\sigma}^{(0,0)}_{bb} = - \Gamma_M \sigma_{bb}^{(0,0)}
+\Gamma_L \sigma_{aa}^{(0,0)}
-\pi\rho_M\Omega_M\bar\Omega_M(\sigma^{(0,0)}_{bc}+\sigma^{(0,0)}_{cb})
\label{a12},
\end{equation}
It corresponds to Eq.~(\ref{a13a}) for $k=n=0$. 
Applying the same procedure to each of equations (\ref{a6c}) we 
obtain the Bloch-type equations, (\ref{a13}) for the density matrix element 
$\sigma^{(k,n)}(t)$.


\begin{references}
\bibitem{foot1}The results are valid also for non-zero temperature providing
that $T\ll E_F^L-E_{1,2},\; E_{1,2}-E_F^R$. Otherwise, the average 
over the reservoir states would enter explicitly in the equations.
\end{references}
\end{document}